\documentclass[twocolumn,secnumarabic,nobibnotes, aps, prl, superscriptaddress,floatfix]{revtex4-1}

\usepackage[normalem]{ulem} 	
\usepackage[usenames,dvipsnames]{color}
\usepackage{upgreek}
\usepackage{braket}
\usepackage{float}
\usepackage[nointegrals]{wasysym}
\usepackage{helvet}
\usepackage{hyperref}
\usepackage{xcolor}
\usepackage[english]{babel}
\usepackage{graphicx, import}
\usepackage{amssymb,amsfonts,amsmath}

\renewcommand{\Im}{\mathop{\rm Im}\nolimits}
\renewcommand{\Re}{\mathop{\rm Re}\nolimits}

\renewcommand{\phi}{\varphi}

\begin{document}

\title{Waveguide bandgap engineering with an array of superconducting qubits }

\author{Jan David Brehm}
	\email[]{jan.brehm@kit.edu}
	\affiliation{Physikalisches Institut, Karlsruhe Institute of Technology, 76131 Karlsruhe, Germany}
\author{Alexander N. Poddubny}
	\affiliation{Ioffe Institute, St. Petersburg 194021, Russia}
\author{Alexander Stehli}
	\affiliation{Physikalisches Institut, Karlsruhe Institute of Technology, 76131 Karlsruhe, Germany}
\author{Tim Wolz}
	\affiliation{Physikalisches Institut, Karlsruhe Institute of Technology, 76131 Karlsruhe, Germany}	
\author{Hannes Rotzinger}
	\affiliation{Physikalisches Institut, Karlsruhe Institute of Technology, 76131 Karlsruhe, Germany}
	\affiliation{Institute for Quantum Materials and Technologies, Karlsruhe Institute of Technology, 76021 Karlsruhe, Germany}
\author{Alexey V. Ustinov}
	\affiliation{Physikalisches Institut, Karlsruhe Institute of Technology, 76131 Karlsruhe, Germany}
	\affiliation{Institute for Quantum Materials and Technologies, Karlsruhe Institute of Technology, 76021 Karlsruhe, Germany}
	\affiliation{National University of Science and Technology MISIS, Moscow 119049, Russia}
	\affiliation{Russian Quantum Center, Skolkovo, Moscow 143025, Russia}

\date{\today}

\begin{abstract}
Waveguide quantum electrodynamics offers a wide range of possibilities to effectively engineer interactions between artificial atoms via a one-dimensional open waveguide. While these interactions have been experimentally studied in the few qubit limit, the collective properties of such systems for larger arrays of qubits in a metamaterial configuration has so far not been addressed.
Here, we experimentally study a metamaterial made of eight superconducting transmon qubits with local frequency control coupled to the mode continuum of a waveguide. By consecutively tuning the qubits to a common resonance frequency we observe the formation of super- and subradiant states, as well as the emergence of a polaritonic bandgap. Making use of the qubits quantum nonlinearity, we demonstrate control over the latter by inducing a transparency window in the bandgap region of the ensemble. The circuit of this work extends experiments with one and two qubits towards a full-blown quantum metamaterial, thus paving the way for large-scale applications in superconducting waveguide quantum electrodynamics.
\end{abstract}
\maketitle 

\section{Introduction}

The recent advances in the field of quantum information processing has led to a rising demand to explore new systems beyond cavity quantum electrodynamics (QED). One promising candidate is waveguide QED, where quantum systems interact coherently with the mode continuum of a waveguide instead of a cavity. After the pioneering works with single qubits, including the demonstration of resonance fluorescence \cite{astafiev_resonance_2010}, quantum correlations of light and single photon routers \cite{hoi_microwave_2013}, attention shifted to the realization of multiple qubits coupled to a common waveguide.
It was derived \cite{lalumiere_input-output_2013} and experimentally verified \cite{loo_photon-mediated_2013} that multiple qubits obtain an infinite range photon mediated effective interaction which can be tuned with the inter-qubit distance $d$. Furthermore, the shared collective excitations are of polaritonic nature with lifetimes ranging from extremely sub- to superradiant relative to the radiative lifetime of the individual qubits \cite{zhang_theory_2019,albrecht_subradiant_2019}. The strong intrinsic nonlinearity of the qubits was recently shown to give rise to partially localized polaritons \cite{zhong_photon-mediated_2020}, topological edge states \cite{ke_radiative_2020,Poshakinskiy2020quantum}, and quantum correlations in the scattered light of the array \cite{fang_one-dimensional_2014}. The collective quantum properties are exploited in the field of quantum metamaterials \cite{Rakhmanov2008,Macha2014}. Here, the quantum coherence of the constituting qubits is used to engineer a global optical response which depends on their quantum state \cite{Asai2015,Ivi2016,Bamba2016,Asai2018}. With respect to quantum information processing, multi-qubit waveguide QED systems could be harnessed in numerous applications such as on demand, highly efficient creation of multi-photon and entangled states \cite{paulisch_generation_2018,gonzalez-tudela_efficient_2017,zhang_heralded_2019-1}, storage devices for microwave pulses \cite{leung_coherent_2012}, atomic mirrors \cite{chang_cavity_2012}, number-resolved photon detection \cite{malz_number-resolving_2019}, slow and even stopped light \cite{everett_stationary_nodate}. Experimentally, waveguide QED systems have been realized on several platforms including atoms \cite{solano_super-radiance_2017}, quantum dots coupled to nanophotonic waveguides \cite{javadi_single-photon_2015} and defect centers in diamonds \cite{sipahigil_integrated_2016}. Even though superconducting qubits feature advantages such as frequency control, high coherence, near perfect extinction and absence of position and number disorder, superconducting multi-qubit waveguide QED systems have been studied only recently to some extend \cite{mirhosseini_cavity_2019, kim_quantum_2020}.\\
Here, we investigate the mode spectrum of a metamaterial formed by a densely spaced array of eight superconducting transmon qubits coupled to a coplanar waveguide. By employing dedicated flux-bias lines for each qubit, we establish control over their transition frequencies. Thus we are able to alter the number of resonant qubits $N$ at will, allowing us to observe super- and subradiant modes as well as the gradual formation of a bandgap. Our spectroscopic measurements show, that through this control the global optical susceptibility of the metamaterial can be tuned. A demonstration for the collective Autler-Townes splitting of 8 qubits is presented, which marks an important step towards the implementation of quantum memories.

\section{Results and discussion}
\subsection{Circuit design and properties}
\begin{figure}[b]
	\includegraphics[width=\columnwidth]{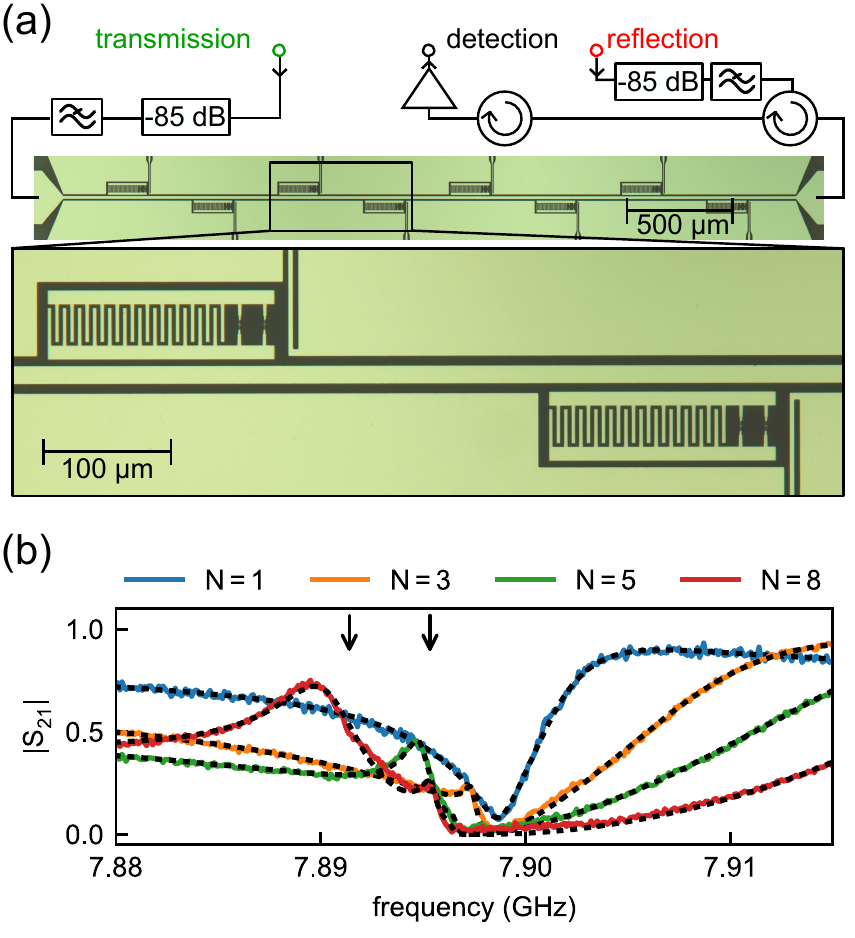}
	\caption{\textbf{Sample and transmission} (a) Optical micrograph of the metamaterial. It is composed of 8 superconducting transmon qubits capacitively coupled to a coplanar waveguide. Local flux-bias lines provide individual qubit frequency control in the range $3-8\,$GHz. (b) Transmission $|S_{21}|$ for different numbers $N$ of resonant qubits at $\omega_\text{r}/2\uppi=7.898\,$GHz and low drive powers. With increasing $N$, the emergence of subradiant states (visible as peaks in transmission) can be observed. Black dotted lines are fits to the expected transmission using a transfer matrix calculation. Black arrows mark calculated frequencies of the two brightest subradiant states for $N=8$.}
	\label{fig1}
\end{figure}
The sample under investigation is depicted in Fig.~\ref{fig1}(a). The spacing $d$ between adjacent qubits is  $400\,\upmu$m, which is smaller than the corresponding wavelength $\lambda$  at all accessible frequencies ($\varphi=\frac{2\uppi}{\lambda}d=0.05-0.16$). The dense spacing is chosen in order to increase the width of the expected bandgap of $\Delta\omega =\Gamma_{10}/\varphi\gg \Gamma_{10}$ and to fulfil the metamaterial limit of subwavelength dimensions \cite{Ivchenko1991}. Here $\Gamma_{10}$ is the radiative decay rate of the individual qubits into the waveguide. The qubits are overcoupled, ensuring a multi-mode Purcell-limited rate $\Gamma_{10}/2\uppi\approx 6.4\,$MHz for high extinction and, simultaneously, better subradiant state visibility.  All qubits are individually frequency controllable between 3 and $8\,$GHz by changing the critical currents of the qubit SQUIDs with local flux-bias lines. We compensate unwanted magnetic crosstalk between the flux-bias lines and neighbouring qubits by extracting and diagonalising the full mutual inductance matrix $M$ (see Methods). This allows us to counteract the parasitic crosstalk flux by sending appropriate currents to all qubits, which are not actively tuned. With this calibration scheme we achieve precise control over the individual qubit frequencies. We estimate the residual crosstalk to be smaller than $10^{-3}$. The effective Hamiltonian of this system, after formally tracing out photonic degrees of freedom and applying the Markov approximation, is described by \cite{albrecht_subradiant_2019},
\begin{equation}
\frac{H_{{\rm eff}}}{\hbar}=\sum_j^{8}(\omega_{j}b_j^\dagger b_j^{\vphantom{\dag}}+ \frac{\chi_j}{2}b_j^\dagger  b_j^\dagger b_j^{\vphantom{\dag}}  b_j^{\vphantom{\dag}})+ {\rm i}\frac{\Gamma_{10}}{2}\sum_{k\ne j}^{8}b_k^\dagger b_j^{\vphantom{\dag}}{\rm e}^{-{\rm i}\frac{\omega d}{c}|k-j|},
\label{hamiltonian}
\end{equation}
with the bosonic creation operator $b_j^\dagger\ket{0}=\ket{{\rm e}_j}$, exciting the $j$-th qubit at frequency $\omega_j$; 
where we assume a $\propto e^{+{\rm i} \omega t}$ time dependence of the excitations.
Here, $\chi_j/2\uppi$ is the qubit anharmonicity for which we find spectroscopically weakly varying values around $-275\,$MHz. The last term of $H_{\rm eff}$ describes the effective qubit-qubit coupling. Due to the specific choice of small $d$ its imaginary part dominates over the real part, leading to a suppressed exchange type interaction between the qubits.  The expected eigenfrequencies $\omega_{\xi}$ of $H_{\rm eff}$ in the single excitation limit and $\chi\rightarrow 0$ are shown in Fig.~\ref{fig2}(a).

\subsection{Bandstructure and collective metamaterial excitations}
\begin{figure}[t]
	\includegraphics[width=\columnwidth]{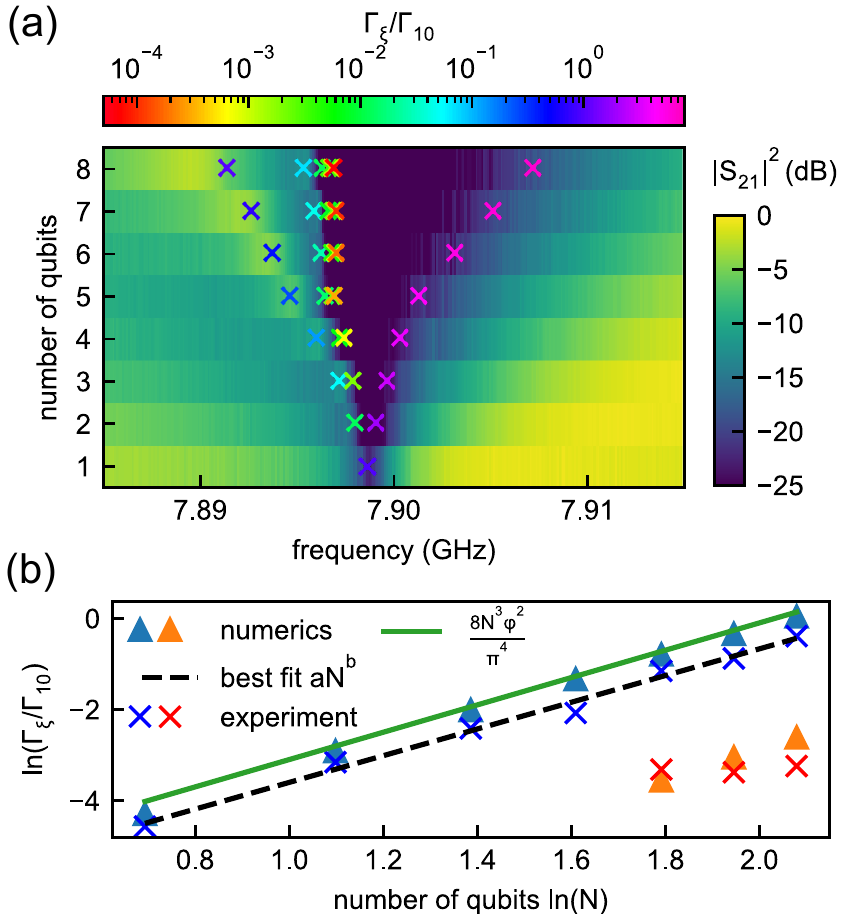}
	\caption{\textbf{Bandgap and polariton relaxation rates} (a) Dependence of absolute transmission $|S_{21}|^2$ on the number of resonant qubits $N$. Crosses mark calculated eigenfrequencies $\omega_{\xi}$ of $H_{\rm eff}$ and their corresponding radiative decay rates $\Gamma_{\xi}/\Gamma_{10}$ (color coded). With increasing $N$ a bandgap of strongly suppressed transmission opens up. (b) Measured radiative decay rates $\Gamma_{\xi}$ of the brightest (blue crosses) and second brightest (red crosses) subradiant states. Orange and blue solid triangles are the corresponding calculated rates, the green line is the analytical result $\Gamma_\xi=8N^3\varphi^2/\uppi^4$. For the brightest subradiant state, a fit to a power-law with exponent $b=2.93\pm0.13$ (black dashed line) confirms the scaling of $\Gamma_\xi\propto N^3$.}
	\label{fig2}
\end{figure}
First, we characterize the mode spectrum of the metamaterial in dependence on the number of resonant qubits $N$ by measuring the transmission coefficient $S_{21}(\omega)$ while the qubits are consecutively tuned to a common resonance frequency at $\omega_\text{r}/2\uppi=7.898\,$GHz, compare Fig.~\ref{fig1} (b) and Fig.~\ref{fig2} (a). The incident photon power  $P_{\rm inc}$  is kept below the single-photon level ($P_{\rm inc}\ll \hbar \omega \Gamma_{10}$) to avoid saturation of the qubits. For a single qubit, the well known resonance-fluorescence was observed as a single dip in transmission \cite{astafiev_resonance_2010}. By fitting the complex transmission data to the expected transmission function $S_{21}(\omega)$ (see Methods) the individual coherence properties of all 8 qubits at $\omega_\text{r}$ can be extracted (see Supplementary Note 1). We find the average radiative rates $\Gamma_{10}/2\uppi=6.4\,$MHz and the intrinsic non-radiative rates $\Gamma_{\rm nr}/2\uppi=240-560\,$kHz.  For $N\ge2$ resonant qubits the system obtains multiple eigenmodes and the super- and subradiant polariton modes start to emerge. The superradiant mode is manifested as a wide transmission dip above $\omega_\text{r}$,  the subradiant modes can be identified  as transmission peaks below $\omega_\text{r}$. We note that the peak shape is created by Fano interference of the sub- and the superradiant modes (see Supplementary Note 5), which also causes the calculated eigenfrequencies of $H_{\rm eff}$ to not exactly coincide with the maximum of the peaks in Fig.~\ref{fig1} (b) and Fig.~\ref{fig2}(a).  For $N\ge 6$ qubits a second darker subradiant mode is visible between $\omega_r$ and the brightest subradiant mode. The limiting factor for the observation of the subradiant states is the intrinsic qubit coherence as given by $\Gamma_{\rm nr}$, setting an upper threshold for the maximum observable lifetime. Darker subradiant states with $\Gamma_{\xi}<\Gamma_{\rm nr}$ decay in the qubits into dielectric channels or dephase due to flux noise in the SQUIDs before they are remitted into the waveguide.
\begin{figure}[t]
	\includegraphics[width=\columnwidth]{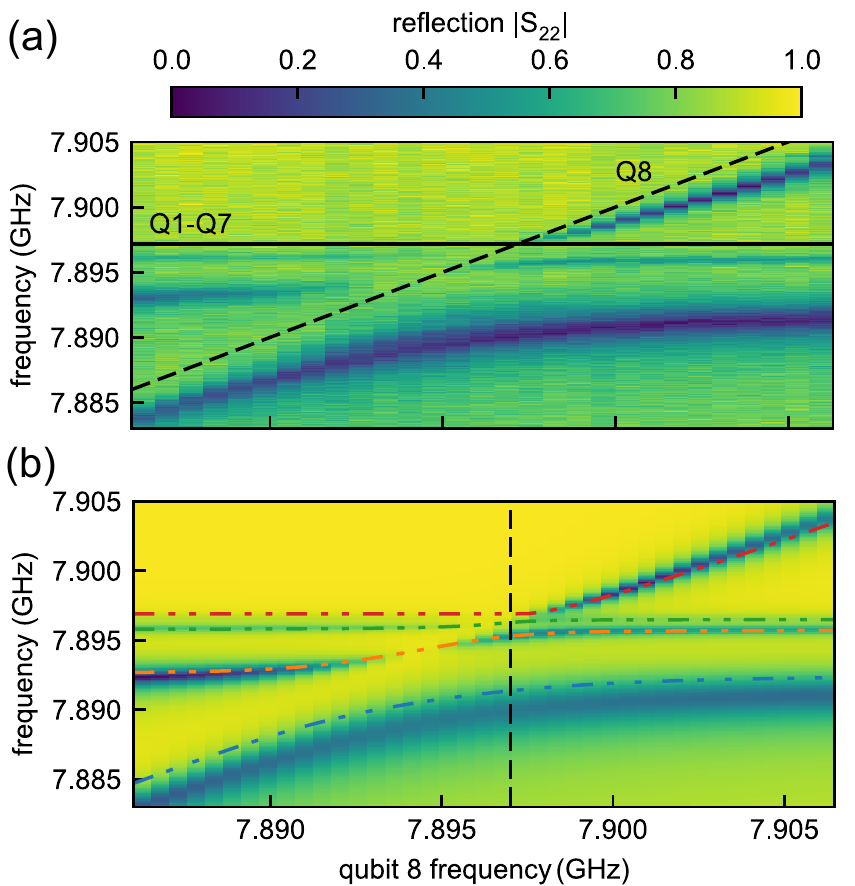}
	\caption{\textbf{Off-resonant qubits} (a) Measured absolute reflection $|S_{22}|$ for $N=7$ resonant qubits at $\omega_\text{r}/2\uppi=7.897\,$GHz (solid black line) and qubit 8 being tuned through the collective resonance (black dashed line). (b) Calculated reflection with transfer matrix method and relevant eigenfrequencies of $H_{\rm eff}$ (dash dotted lines) show good agreement with experimental result in (a). The vertical line indicates the resonance with $\omega_8=\omega_\text{r}$,  corresponding to the situation observed in Fig.~\ref{fig1}.
	}
	\label{fig3}
\end{figure} 
The observed transmission coefficient $|S_{21}|$ is in good agreement with the calculated transmission based on a transfer matrix approach \cite{asenjo-garcia_exponential_2017}. The asymmetric lineshape of the resonances is a parasitic effect caused by interference of the signal with low-Q standing waves in the cryostat \cite{khalil_analysis_2012}. We account for this effect in the transfer matrix calculation by adding semi-reflective inductances in front and after the qubit array. As shown in Fig.~\ref{fig2}(a), a frequency region  of strongly suppressed transmission with $|S_{21}|^2<-25\,$dB is opening up above $\omega_\text{r}$ with increasing $N$. This effect is associated with the emergence of a polaritonic bandgap, where the effective refractive index becomes purely imaginary, as expected for any kind of resonant periodic structures \cite{Ivchenko1991,tsoi_quantum_2008}. For $N=8$ qubits we extract a bandgap bandwidth of $\Delta\omega\approx 1.9\Gamma_{10}$. Compared to the expected bandgap width of $\Delta\omega=\Gamma_{10}/\varphi\approx 6.3\Gamma_{10}$ of the structure for $N\rightarrow\infty$ this places our system size in the transitioning regime between a single atom and a fully extended metamaterial with a continuous mode spectrum. The radiative decay rates $\Gamma_\xi$ of the observed subradiant states shown in Fig.~\ref{fig2}(b) are extracted by fitting Lorentzians to the corresponding modes in the reflection data (not shown). It can be generally shown, that $|S_{21}|^2$ indeed obtains a Lorentzian shape in the vicinity of each $\omega_\xi$ \cite{Ivchenko1994,Kosobukin2007,asenjo-garcia_exponential_2017}. From a fit to a power law $\propto N^b$ with exponent $b=2.93\pm0.13$ we find that the rate of the brightest subradiant states scales with $\Gamma_\xi\propto N^3$, which we also find analytically for densely packed qubit structures with $\varphi\ll1$ from $H_{\rm eff}$ (see Supplementary Note 4). The found scaling is the complementary asymptotic of the theoretically predicted $\Gamma_\xi\propto N^{-3}$ law for the darkest subradiant modes in references \cite{albrecht_subradiant_2019,zhang_theory_2019,tsoi_quantum_2008}. Small deviations between calculated and measured values of $\Gamma_{\xi}$ are caused by imperfect qubit tuning and distortions of the observed reflection coefficient due to the microwave background. 
\begin{figure*}[t!]
	\includegraphics[]{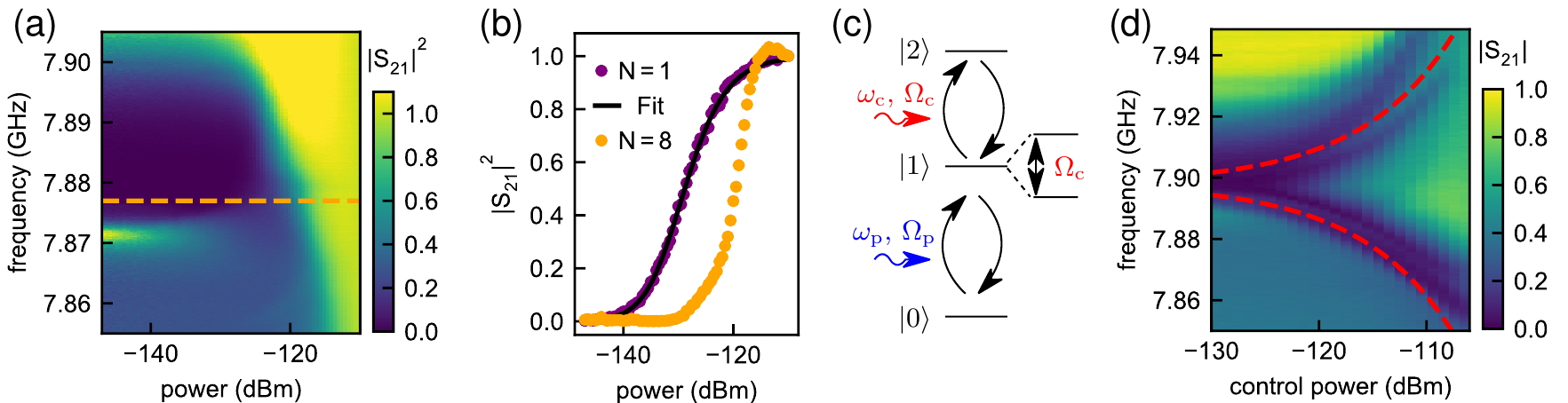}
	\caption{\textbf{Power saturation and Autler-Townes splitting} (a) Absolute power transmission $|S_{21}|^2$ for $N=8$ resonant qubits shows saturation with increasing power due to the anharmonic level structure of the transmons. (b) Comparison between the saturation of $|S_{21}|^2$ at $\omega_\text{r}$ for a single qubit and $N=8$ qubits. (c) Schematic illustration of the Autler-Townes effect for a ladder-type three-level system. A control tone is driving the $1\rightarrow 2$ transition with Rabi strength $\Omega_\text{c}$ and frequency $\omega_{\rm c}$, while a weak tone is probing the transmission of the $0\rightarrow 1$ transition. (d) Experimental demonstration of the collective Autler-Townes splitting for $N=8$. The red dashed line marks the fitted level separation to $ \Omega_{\rm c}$.
	}
	\label{fig4}
\end{figure*}
The control over the mode spectrum can be further elaborated with the off-resonant situation, where one qubit has a finite detuning $\Delta$ from the common resonance frequency $\omega_\text{r}$ of the residual qubits by sweeping it through the common resonance as shown in Fig.~\ref{fig3}(a). Here, qubit 8 is tuned through the collective resonance of qubits 1-7. For large detunings $|\Delta|\gg \Gamma_{10}$ the modes of the ensemble and qubit 8 are not hybridized (not shown). For smaller detunings an additional partially hybridized subradiant mode appears, which becomes for $\Delta\approx 0$ the brightest subradiant state of the fully hybridized 8-qubit system. The result is in good agreement with the transfer matrix calculation and direct diagonalization of $H_{\rm eff}$ as shown in Fig.~\ref{fig3}(b). The level repulsion between the eigenstates is caused by the residual exchange-type interaction between the qubits due to the finite inter-qubit distance $d$. In Fig.~\ref{fig3} there are several blind spots where the subradiant states turn completely dark, occurring when the frequency of the detuned qubit matches the frequency of a dark mode. 
This is explained by the Fano-like interferences \cite{Limonov2017} between the detuned qubit resonance and the modes of the resonant qubits, which are analyzed in more detail in the Supplementary Note 5.

\subsection{Collective Autler-Townes splitting}
The intrinsic quantum nonlinearity of the system due to the anharmonic nature of the qubits can be probed by increasing the microwave power beyond the single photon regime ($P_{\rm inc}> \hbar \omega \Gamma_{10}$). At higher power the $0\rightarrow 1$ transition of a single qubit will start to saturate and transmission at resonance will increase back to unity \cite{astafiev_resonance_2010}, see Fig.~\ref{fig4}(b).
We could experimentally verify the prediction of the saturation of an ensemble of resonant qubits at higher drive rates 
of Refs.~\cite{lalumiere_input-output_2013, astafiev_resonance_2010}, which can be observed in Fig.~\ref{fig4}(a). We point out that all spectroscopic features of the metamaterial, such as super- and subradiant modes as well as the bandgap, are saturable. The power $P_{50\%}$ needed to saturate the transmission at $\omega_{r}$ to 50\% ($|S_{21}|^2=0.5$) grows with an approximate scaling $\propto\ln(N)$. The observed quantum nonlinear behavior establishes a border to previously studied metamaterials consisting of harmonic resonators rather than qubits \cite{mirhosseini_superconducting_2018}. Furthermore the anharmonic level structure of the transmon can be used to electromagnetically open a transparency window around the frequency of the $0\rightarrow 1$ transition, by employing the Autler-Townes splitting (ATS) \cite{abdumalikov_electromagnetically_2010}. As indicated in Fig.~\ref{fig4}(c), a coherent control tone with Rabi strength $\Omega_{\rm c}$ and frequency $\omega_{\rm c}$ drives the $1\rightarrow 2$ transition, while a weak microwave tone with $\Omega_{\rm p}\ll\Omega_{\rm c}$ is probing the transmission at frequencies $\omega_{\rm p}$ around the $0\rightarrow 1$ transition. The control tone is dressing the first qubit level, creating two hybridized levels, which are separated proportionally to its amplitude, and is therefore creating a transparency window with respect to the probe. To the best of our knowledge this effect was so far only demonstrated for a single qubit in superconducting waveguide QED, leading to applications like single photon routers \cite{hoi_microwave_2013}. Here, we observe the collective ATS for up to $N=8$ resonant qubits (Fig.~\ref{fig4}(d)). Analogously to the single qubit case, the observed level splitting is proportional to $\Omega_{\rm c}$. At control tone powers $>-110\,$dBm the bandgap is rendered fully transparent and transmission close to unity is restored. In the experiment we find slightly smaller than unity values of $|S_{21}|\approx 0.75$ due to the interference of the signal with the microwave background. In the bandstructure picture the collective ATS can be understood as the dressed states of the individual qubits giving rise to two independent Bloch bands \cite{witthaut_photon_2010}. Therefore, the two branches of the collective ATS with suppressed transmission are collectively broadened bandgaps and have a larger linewidth than the dressed states of the single qubit ATS. As pointed out in reference \cite{witthaut_photon_2010} the collective ATS demonstrates active control over the bandstructure of the metamaterial via the parameters $\omega_\text{c}$ and $\Omega_\text{c}$. The observed splittings are in agreement with a transfer matrix calculation and a full master equation simulation (see Supplementary Note 6). Minor deviations are caused by imperfect qubit tuning and differing qubit anharmonicities of about $\sigma_\chi\approx 10\,$MHz. We argue here that the single photon router concept with a single qubit can significantly be improved with multiple qubits, which form a much wider stop-band with higher saturation power, thus permitting to route also multiple photons.

In conclusion, we demonstrated a fully controllable quantum metamaterial consisting of 8 densely packed transmon qubits coupled to a waveguide. Such intermediate system size combined with individual qubit control allowed us to explore the transition from a single mode regime to a continuous band spectrum. By tuning the qubits consecutively to resonance, we observed the emergence of a polaritonic  bandgap, and confirmed the scaling of the brightest dark mode decay rate with the qubit number. Active control over the band structure of the ensemble was demonstrated by inducing a transparency window in the bandgap region, using the Autler-Townes effect. Our work promotes further research with higher qubit numbers to realize a long-living quantum memory.

\section{Methods}
\subsection{Fabrication}
The sample is fabricated with two consecutive lithography steps from thermally evaporated aluminum in a Plassys MEB550s shadow evaporator on a 500 $\upmu$m sapphire substrate. In a first step solely the qubits are patterned with $50\,$keV electron beam lithography. The Josephson junctions are patterned with a bridge-free fabrication technique \cite{lecocq_junction_2011} using a PMMA/PMMA-MAA double resist stack. Before double-angle evaporation, the developed resist stack is cleaned for 6 min. with an oxygen-plasma to remove resist residuals in the junction area to reduce the impact of junction aging \cite{lecocq_junction_2011}. In a second optical lithography step we pattern the CPW-waveguide and the ground-plane in a liftoff-process on S1805 photo resist. The SQUIDs are formed by two Josephson junctions, enclosing an area of 560$\,\upmu$m$^2$. By design, the junction areas are 0.12 and 0.17$\,\upmu$m$^2$ with a designed asymmetry of $17\,$\%.
\subsection{Calibration of magnetic crosstalk}
We extract the full 8x8 mutual inductance matrix $M$ between the qubits and the bias coils in 28 consecutive measurements. For that, the transmission through the chip is observed at a fixed frequency while tuning the currents of two qubits $I_x$ and $I_y$ such that they get tuned through the observation frequency. Fitting the slope of the observed qubit traces gives access to $M_{xy}/M_{xx}$ and $M_{yx}/M_{yy}$. When $M$ is known, compensation currents which are send to all qubits which are not actively tuned, can be calculated to compensate unwanted crosstalk. We estimate the residual crosstalk to be smaller than $0.1\,\%$. Further information is provided in the Supplementary Note 2.
\subsection{Normalization of spectroscopic data}
We normalize the transmission data by dividing the raw data $S^\text{raw}_{21}$ by the transmission data at high powers $S^\text{sat}_{21}$: $S^\text{calib}_{21}(\omega)=S^\text{raw}_{21}(\omega)/(a S^\text{sat}_{21}(\omega))$, where $a\approx1$ is a constant factor accounting for weak fluctuations of the amplifier gain. Reflection data is normalized by dividing the raw data $S^\text{raw}_{22}$ with its maximum value at the qubit resonance frequency $f_\text{r}$: $S^\text{calib}_{22}=S^\text{raw}_{22}/S^\text{raw}_{22}(f_\text{r})$. This approximation is justified for the sample under investigation since the extinction of single and multiple qubits is very close to 1. A more rigorous normalization of the data based on energy conservation $|S_{21}|^2+|S_{11}|^2\approx 1$ is not applicable here, due to differing signal paths for reflection and transmission measurements, as pointed out in the main text. Additionally, in any experimental system the insertion losses due to impedance mismatches of the feedline can not be avoided. Since they are in general not symmetric on both sides of the sample, $|S_{11}|\neq|S_{22}|$ and therefore $|S_{21}|^2+|S_{22}|^2\neq 1$ .
\subsection{Characterization of individual qubits}
The amplitude transmission coefficient  of a driven two-level system side-coupled to a waveguide $S_{21}$ is given by \cite{hoi_microwave_2013}:
\begin{equation}
S_{21}=1-\frac{\Gamma_{10}}{2\gamma_{10}}\frac{1-{\rm i}\frac{\omega-\omega_\text{r}}{\gamma_{10}}}{1+\left(\frac{\omega-\omega_\text{r}}{\gamma_{10}}\right)^2+\frac{\Omega_{\rm p}^2}{(\Gamma_{10}+\Gamma_l)\gamma_{10}}}
\label{2L}
\end{equation}
The decoherence rate $\gamma_{10}=\Gamma_{10}/2+\Gamma_{\rm nr}$ is the sum of radiative decay $\Gamma_{10}$ and non-radiative decay rates $\Gamma_{\rm nr}=\Gamma_{\Phi}+\Gamma_l/2$. Here $\Gamma_{\Phi}$ accounts for pure dephasing of the qubit and $\Gamma_l$ for all non radiative relaxation channels. We define the extinction coefficient as $1-(1-\Gamma_{10}/2\gamma_{10})^2$, measuring the suppression of power-transmission at very low drive powers. A circle fitting procedure \cite{probst_efficient_2015} is used to fit equation \ref{2L} to the measured complex transmitted signal $S_{21}$ in the limit of weak driving $\Omega_{\rm p}^2\ll(\Gamma_{10}+\Gamma_l)\gamma_{10}$. The decoherence rates of the individual qubits and further details on the fitting procedure are provided in the Supplementary Note 1.

\section{Acknowledgement}
This work has received funding from the Deutsche Forschungsgemeinschaft (DFG) by the Grant No. US 18/15-1, the European Union’s Horizon 2020 Research and Innovation Programme under Grant Agreement No. 863313 (SUPERGALAX), and by the Initiative and Networking Fund of the Helmholtz Association, within the Helmholtz Future Project ‘Scalable solid state quantum computing’. We acknowledge financial support from Studienstiftung des Deutschen Volkes (JDB),  Landesgraduiertenf\"orderung-Karlsruhe (AS) and Helmholtz International Research School for Teratronics (TW). Analysis of subradiant mode lifetimes, performed by ANP, has been supported by the Russian Science Foundation Grant 20-12-00194. Basic concepts for this work were developed with the financial support from the Russian Science Foundation (contract No. 16-12-00095). AVU acknowledges partial support from the Ministry of Education and Science of the Russian Federation in the framework of the Increase Competitiveness Program of the National University of Science and Technology MISIS (contract No. K2-2020-022).

\section{Author contributions}
J.D.B. fabricated the devices supported by H.R.. J.D.B. performed the measurements with support of A.S. and T.W.. A.S. developed concepts for the crosstalk calibration. A.V.U. and H.R. setup the measurement facility. J.D.B. analyzed the data. A.N.P. performed the calculations on Fano-interference and linewidth-scaling. J.D.B. and A.N.P. wrote the paper. All authors contributed to the discussion. The project was supervised by H.R. and A.V.U.

\bibliography{BandgapEngineering_bib}

\clearpage
\onecolumngrid

\appendix

\setcounter{figure}{0}
\setcounter{equation}{0}
\renewcommand{\figurename}{Supplementary Figure}
\renewcommand{\tablename}{Supplementary Table}
\renewcommand{\thetable}{\arabic{table}}

\section{Supplementary Note 1: Qubit characterization}

\begin{figure}[htb!]
	\includegraphics[width=\columnwidth]{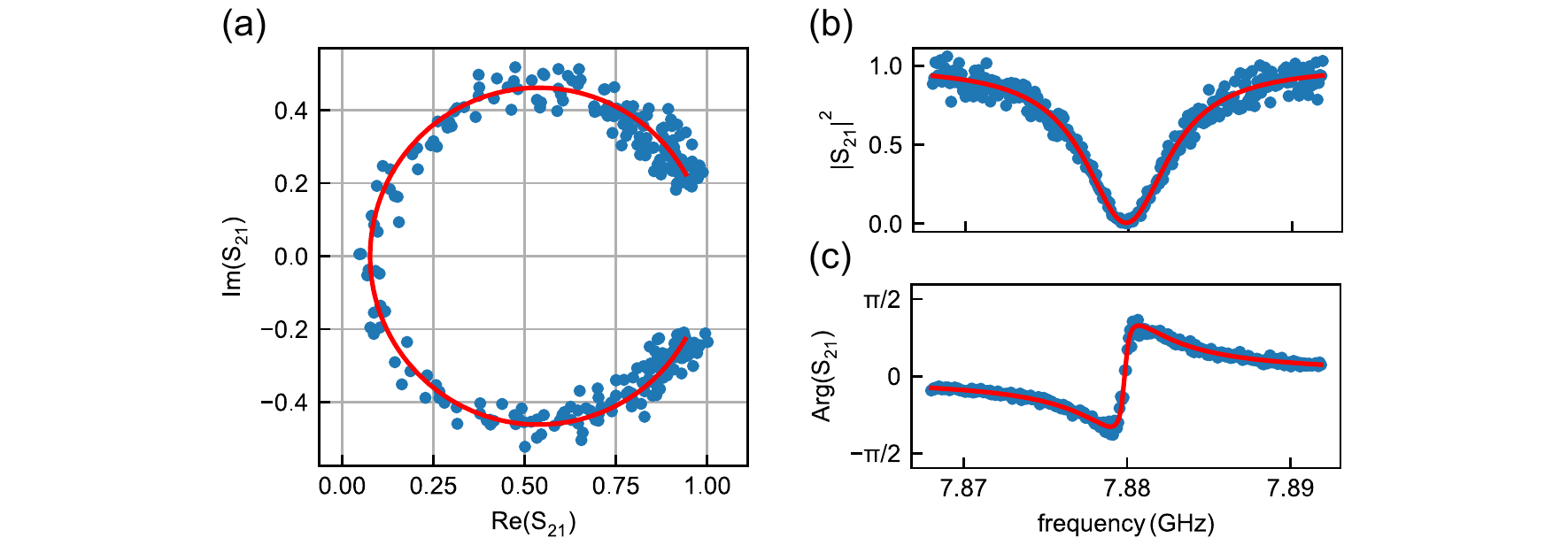}
	\caption{(a) Complex $S_{21}$ transmission signal of qubit 1 at $7.88\,$GHz with normalized background at low power. Red line is circle fitted theory (Eq.~(\ref{2L})). Transmission signal (b) amplitude $|S_{21}|^2$, and (c) phase signal $\text{Arg}(S_{21})$.
	}
	\label{circlefit}
\end{figure}

The amplitude reflection coefficient  of a driven two-level system side-coupled to a waveguide $r$ is given by \cite{hoi_microwave_2013}
\begin{equation}
	r=-\frac{\Gamma_{10}}{2\gamma_{10}}\frac{1-{\rm i}\frac{\omega-\omega_\text{r}}{\gamma_{10}}}{1+\left(\frac{\omega-\omega_\text{r}}{\gamma_{10}}\right)^2+\frac{\Omega_{\rm p}^2}{(\Gamma_{10}+\Gamma_l)\gamma_{10}}}\:,
	\label{2L}
\end{equation}
where 
$\omega_{10}$ is the resonance frequency of the 0-1 transmon transition. The decoherence rate $\gamma_{10}=\Gamma_{10}/2+\Gamma_{\rm nr}$ is the sum of radiative decay $\Gamma_{10}$ and non-radiative decay rates $\Gamma_{\rm nr}=\Gamma_{\Phi}+\Gamma_l/2$. Here $\Gamma_{\Phi}$ accounts for pure dephasing of the qubit and $\Gamma_l$ for all non radiative relaxation channels. In the approximation that the non-radiative relaxation is negligible compared to the pure dephasing rate, we can assign the typical coherence times $T_1=1/\Gamma_{10}$, $T_2=1/\gamma_{10}$, and $T_\Phi=1/\Gamma_{\rm nr}$  to these rates. The approximation is justified since here $\Gamma_{\rm nr}$ is strongly dominated by flux noise in the SQUID for frequencies detuned from the flux-sweetspot. We define the extinction coefficient as $1-(1-\Gamma_{10}/2\gamma_{10})^2$, measuring the suppression of power-transmission at very low drive powers. The complex transmission is given by $t=1+r$.\\
In order to extract the lifetimes of the individual qubits we use a circle fitting procedure \cite{probst_efficient_2015} to fit Eq.~(\ref{2L}) to the measured complex transmitted signal in the limit of weak driving $\Omega_{\rm p}^2\ll(\Gamma_{10}+\Gamma_l)\gamma_{10}$. Supplementary Figure~\ref{circlefit} depicts an exemplary fit to the complex transmission data of qubit 1 at $7.88\,$GHz, showing the resonance-florescence.
Due to the local dc flux-bias lines, all qubits are frequency tunable between their upper and lower sweetspot, ranging from 3 to 8 GHz. By fitting the SQUID-dispersion we extract the exact sweetspot positions ($f^\text{min}_{01}$, $f^\text{max}_{01}$) in Supplementary Table~\ref{singlequbitparamsgeneral}. Furthermore, the qubit anharmonicities $\chi$ around the upper sweet spot were extracted by measuring the Autler-Townes splitting of all qubits. We note that the data presented in this work was measured in several consecutive cooldowns, leading to slightly changed sweetspot positions of around $20\,$MHz due to junction aging. Supplementary Table~\ref{singlequbitparamsgeneral} shows the extracted lifetimes around $7.9\,$GHz, the frequency of the lowest upper sweetspot and thus the most favourable position for the experiments in the main text.\\

\begin{table}
	\caption{Measured individual qubit properties around $7.9\,$GHz and upper- and lower-sweetspot positions $f^{min}_{01}$, $f^{max}_{01}$.}
	\begin{ruledtabular}
		\begin{tabular}{c| c c c c c c c c}
			Parameter & Qubit 1 & Qubit 2 & Qubit 3 & Qubit 4 & Qubit 5 & Qubit 6 & Qubit 7 & Qubit 8\\
			$T_1\,$(ns)&27.6& 22.5&  22.4&  20.0& 22.0& 27.6& 28.0&  34.3\\
			
			$T_2\,$(ns)&51.0 &  41.4&  41.4&  35.7&	39.9& 47.7&  47.9& 55.3\\
			
			$T_\Phi\,$(ns)&665.1&  520.0&  557.6&  334.6&			422.9& 351.1&  329.0& 284.0 \\
			
			Ext. Coeff$\,\%$&99.4&  99.4&  99.4&  98.7&	99.1&98.2&  97.9& 96.2\\

			$|\chi|/2\uppi\,(\text{MHz})$&283&279&273&275&267&281&273&276\\
			$f^{max}_{01}\,(\text{GHz})$&8.097&7.900&8.088&8.114&8.115&7.95&8.066&8.136\\ 
			$f^{min}_{01}\,(\text{GHz})$&3.029&3.091&2.912&2.986&2.970&2.936&2.588&2.484\\
		\end{tabular}
	\end{ruledtabular}
	\label{singlequbitparamsgeneral}
\end{table}

\section{Supplementary Note 2: Magnetic crosstalk calibration}
The fluxes ($\Phi_1 ... \Phi_8$) in all SQUIDs are related with the applied currents via the mutual inductance matrix \textbf{M}:
\begin{equation}
	\begin{pmatrix}
		\Phi_1\\
		\Phi_2\\
		\Phi_3\\
		\Phi_4\\
		\Phi_5\\
		\Phi_6\\
		\Phi_7\\
		\Phi_8\\
	\end{pmatrix}=
	\begin{pmatrix}
		M_{11} & M_{12} &M_{13} &M_{14} &M_{15} &M_{16} &M_{17} &M_{18} \\
		M_{21} & M_{22} &M_{23} &M_{24} &M_{25} &M_{26} &M_{27} &M_{28} \\
		M_{31} & M_{32} &M_{33} &M_{34} &M_{35} &M_{36} &M_{37} &M_{38} \\
		M_{41} & M_{42} &M_{43} &M_{44} &M_{45} &M_{46} &M_{47} &M_{48} \\
		M_{51} & M_{52} &M_{53} &M_{54} &M_{55} &M_{56} &M_{57} &M_{58} \\
		M_{61} & M_{62} &M_{63} &M_{64} &M_{65} &M_{66} &M_{67} &M_{68} \\
		M_{71} & M_{72} &M_{73} &M_{74} &M_{75} &M_{76} &M_{77} &M_{78} \\
		M_{81} & M_{82} &M_{83} &M_{84} &M_{85} &M_{86} &M_{87} &M_{88} \\
	\end{pmatrix}
	\begin{pmatrix}
		I_1\\
		I_2\\
		I_3\\
		I_4\\
		I_5\\
		I_6\\
		I_7\\
		I_8\\
	\end{pmatrix}
\end{equation}
\begin{figure}[htb!]
	\includegraphics[width=\columnwidth]{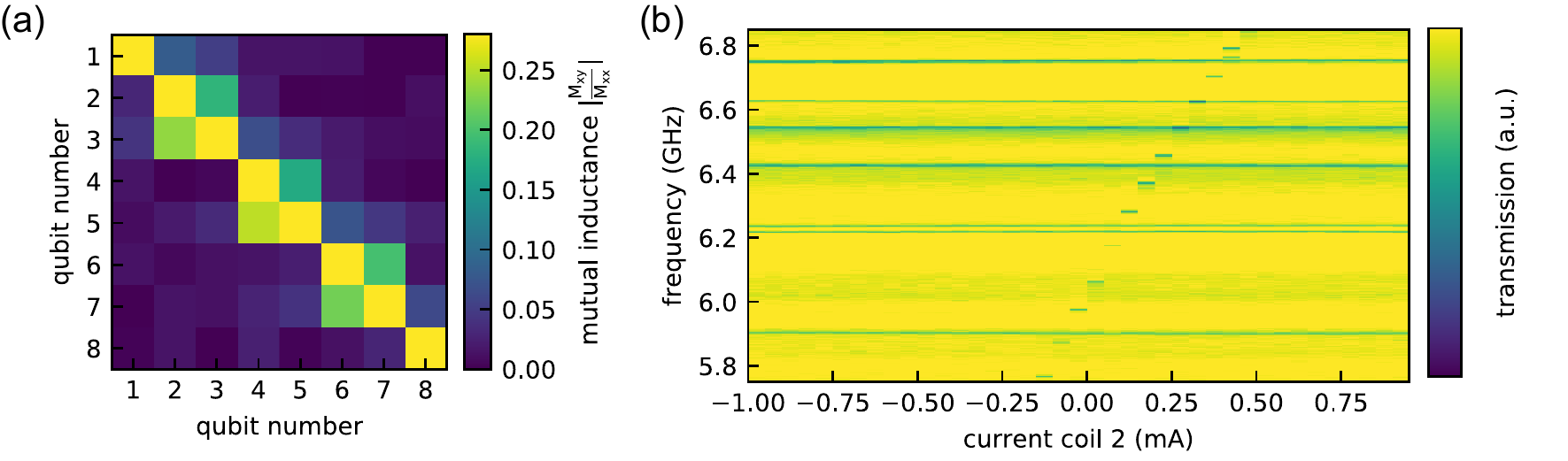}
	\caption{(a) Mutual inductance matrix between the bias coils. (b) Transmission measurement with qubit 2 being tuned, while crosstalk calibration is used. Horizontal lines are the other 7 qubits, not changing their frequency.
	}
	\label{crosstalkmatr}
\end{figure}

The mutual inductance matrix elements can be extracted by observing the transmission through the 8 qubit chip at a fixed frequency while simultaneously tuning the frequency of two qubits $x$ and $y$ through this observation frequency by tuning the currents $I_x$ and $I_y$.
The observation frequency is chosen such that the qubits have a steep flux-dispersion. We assume that the qubit frequencies are proportional to the flux in their coils, which is satisfied for not too large frequency changes. From the equation above we get:
\begin{eqnarray}
	\text{const}&=&\Phi_x=M_{xx}I_x +M_{xy} I_y \qquad \rightarrow \qquad I_y=-\frac{M_{xx}}{M_{xy}}I_x+\frac{\Phi_y}{M_{xy}}\\
	\text{const}&=&\Phi_y=M_{yy}I_y +M_{yx} I_x \qquad \rightarrow \qquad I_x=-\frac{M_{yy}}{M_{yx}}I_y+\frac{\Phi_x}{M_{yx}}
\end{eqnarray}
Therefore, by fitting the slopes of the two qubit lines visible in this measurement, the mutual inductance matrix elements $\frac{M_{xy}}{M_{xx}}$ and $\frac{M_{yx}}{M_{yy}}$ can be extracted. In case of an 8-qubit chip all 28 possible combinations between two coils have to be measured and fitted. The extracted mutual inductance matrix (with each element normalized to the diagonal element of its line) for the 8-qubit chip is shown  in Supplementary Figure~\ref{crosstalkmatr}(a). The figure shows that only the nearest neighbor coupling goes beyond $10\,$\% of the self-inductance. The crosstalk is only large for every second pair of neighbors, due to the specific placement of the on-chip bond-wires on this sample.\\

As soon as the mutual inductance matrix is known the crosstalk can be compensated by setting a compensation current to all seven other coils while one qubit is effectively tuned. If, for example, qubit 4 is effectively tuned, the compensation currents for all other coils can be calculated by solving the following system of linear equations:
\begin{equation}
	\begin{pmatrix}
		0\\
		0\\
		0\\
		0\\
		0\\
		0\\
		0\\
	\end{pmatrix}\overset{!}{=}
	\begin{pmatrix}
		\Phi_1\\
		\Phi_2\\
		\Phi_3\\
		\Phi_5\\
		\Phi_6\\
		\Phi_7\\
		\Phi_8\\
	\end{pmatrix}=
	\begin{pmatrix}
		M_{11} & M_{12} &M_{13} &M_{14} &M_{15} &M_{16} &M_{17} &M_{18} \\
		M_{21} & M_{22} &M_{23} &M_{24} &M_{25} &M_{26} &M_{27} &M_{28} \\
		M_{31} & M_{32} &M_{33} &M_{34} &M_{35} &M_{36} &M_{37} &M_{38} \\
		M_{51} & M_{52} &M_{53} &M_{54} &M_{55} &M_{56} &M_{57} &M_{58} \\
		M_{61} & M_{62} &M_{63} &M_{64} &M_{65} &M_{66} &M_{67} &M_{68} \\
		M_{71} & M_{72} &M_{73} &M_{74} &M_{75} &M_{76} &M_{77} &M_{78} \\
		M_{81} & M_{82} &M_{83} &M_{84} &M_{85} &M_{86} &M_{87} &M_{88} \\
	\end{pmatrix}
	\begin{pmatrix}
		I^\text{comp}_1\\
		I^\text{comp}_2\\
		I^\text{comp}_3\\
		I_4\\
		I^\text{comp}_5\\
		I^\text{comp}_6\\
		I^\text{comp}_7\\
		I^\text{comp}_8\\
	\end{pmatrix}
\end{equation}
Supplementary Figure~\ref{crosstalkmatr}(b) shows the applied compensation procedure while only qubit 2 is effectively tuned. We estimate the residual crosstalk to be below $1\,\permil$.
\section{Supplementary Note 3: Transfer matrix approach}
\begin{figure}[htb!]
	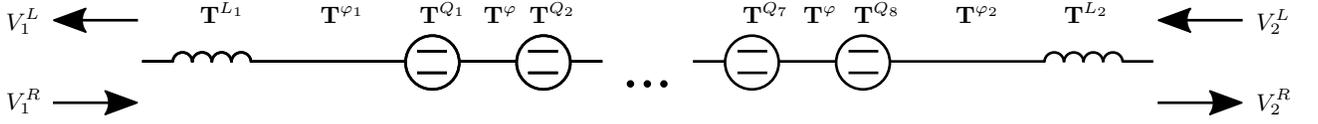
	\caption{Schematic representation of the transfer matrix
	}
	\label{tmatrix}
\end{figure}
The transfer matrix connects the ingoing and outgoing field amplitudes of a two-port network in the following way \cite{asenjo-garcia_exponential_2017}:
\begin{equation}
	\begin{pmatrix}
		V_2^R\\V_2^L
	\end{pmatrix}
	=\text{\textbf{T}}
	\begin{pmatrix}
		V_1^R\\V_1^L
	\end{pmatrix}
\end{equation}
For a combined network of multiple subsystems in series, the total transfer matrix $\text{\textbf{T}}_\text{tot}$ is the product of the individual transfer matrices.  The transmission- $S_{21}$ and reflection-coefficient $S_{22}$ can be recovered from $\text{\textbf{T}}_\text{tot}$ with the following relations (assuming $V_2^L=0$ for transmission experiment and $V_1^L=0$ for reflection experiment):
\begin{equation}
	\frac{V_2^R}{V_1^R}=\frac{1}{T_{22}^\text{tot}}=S_{21}\qquad \text{and} \qquad \frac{V_2^R}{V_2^L}=\frac{T_{12}^\text{tot}}{T_{11}^\text{tot}}=S_{22}
\end{equation}
The full system under consideration is depicted in Supplementary Figure~\ref{tmatrix}. To be able to account also for the asymmetric line shape as encountered in the experiment, we include two inductances $L_{1,2}$  on the edges of the system, acting as semi-transparent mirrors. The total transfer matrix then reads:
\begin{equation}
	\text{\textbf{T}}_\text{tot}=  \text{\textbf{T}}^{L_2}\text{\textbf{T}}^{\phi_2}\text{\textbf{T}}^{Q_8}\text{\textbf{T}}^\phi...\text{\textbf{T}}^\phi\text{\textbf{T}}^{Q_2}\text{\textbf{T}}^\phi\text{\textbf{T}}^{Q_1}\text{\textbf{T}}^{\phi_1}\text{\textbf{T}}^{L_1}
\end{equation}
$\text{\textbf{T}}^{Q_n}$ is the transfer matrix of the n-th qubit and is given by:
\begin{equation}
	\text{\textbf{T}}^{Q_n}=\begin{pmatrix}
		\frac{1+2r}{1+r}&\frac{r}{1+r}\\
		-\frac{r}{1+r}&\frac{1}{1+r}
	\end{pmatrix}
	\label{tmatrixdef}
\end{equation}
with $r$ being the reflection coefficient of the qubit, Eq.~(\ref{2L}).
In order to reduce the number of free parameters in the fitting procedure, we assume that the reflection coefficient $r$ of all qubits is identical. $\text{\textbf{T}}^\phi$ accounts for the propagation along a bare piece of transmission line: 
\begin{equation}
	\text{\textbf{T}}^\phi=\begin{pmatrix}
		\exp{(-{\rm i}\phi)}&0\\
		0&\exp{({\rm i}\phi)}
	\end{pmatrix}
\end{equation}
, where the propagating field accumulates a phase of $\phi=kd=\frac{\omega}{c}d$, with length $d$ and phase velocity $c$. For our specific choice of Sapphire substrate and a waveguide in cpw-geometry (gap: $7.7\,\upmu$m, center: $14.4\upmu$m) we obtain $c=1.2\cdot10^8\,$m/s.\\
$\text{\textbf{T}}^L$ is the transfer matrix of an inductor creating an impedance mismatch compared to the $Z_0=50\,\Omega$ environment:
\begin{equation}
	\text{\textbf{T}}^L=\begin{pmatrix}
		1-\frac{{\rm i}\omega L}{2Z_0}&-\frac{{\rm i}\omega L}{2Z_0}\\
		\frac{{\rm i}\omega L}{2Z_0}&1+\frac{{\rm i}\omega L}{2Z_0}
	\end{pmatrix}
\end{equation}
\section{Supplementary Note 4: Radiative linewidth scaling of subradiant modes with $N$}
In order to extract the lifetimes of the polaritonic modes in the single excitation sector, we use only the last term of $H_{\rm eff}$ in the main text:
\begin{equation}
	H_\text{rs}={\rm i}\frac{\Gamma_{10}}{2}\exp(-{\rm i}\varphi|r-s|)
	\label{interactham}
\end{equation}
with $H_\text{rs}$ being the interaction matrix element between qubit $r$ and $s$. $\varphi=\frac{\omega_\text{r}}{c}d$ is the phase between two neighbouring qubits.
The inverse Hamiltonian of (\ref{interactham}) $[H^{-1}]_{\text{rs}}$ is exactly 3-diagonal \cite{Poddubny2019quasiflat}:
\begin{equation}\label{eq:matrix}
	[H^{-1}]_{\text{rs}}=\frac{2}{\Gamma_{10}}\begin{pmatrix}
		-\frac{1}{2}\cot{\varphi}-\frac{{\rm i}}{2} & \frac{1}{2\sin{\varphi}} & 0 & \cdots \\
		\frac{1}{2\sin{\varphi}}& -\cot{\varphi}&\frac{1}{2\sin{\varphi}} & \cdots \\
		& \ddots & \ddots&  \\
		\cdots & \frac{1}{2\sin{\varphi}}& -\cot{\varphi}&\frac{1}{2\sin{\varphi}}\\
		\cdots  & 0 &  \frac{1}{2\sin{\varphi}} & -\frac{1}{2}\cot{\varphi}-\frac{{\rm i}}{2} \\
	\end{pmatrix}
\end{equation}	
This means that the Schr\"odinger equation
\begin{equation}\label{eq:iS}
	H^{-1}\psi_\xi=\frac{1}{\omega_{\xi}}\psi_\xi
\end{equation}
of the inverse Hamiltonian is just a tight-binding model. The radiative decay due to the photon escape into the waveguide is present only at the edges of the qubit array and can be treated as a perturbation. For the lower polariton branch we obtain \cite{zhang_theory_2019,vladimirova1998}:
\begin{equation}\label{eq:ans}
	\frac{{\Gamma_{10}}/2}{\omega_{\xi}}=-\frac{2}{\varphi}\sin^2\frac{k}{2}-\frac{2{\rm i}}{N}\cos^2\frac{k}{2},\qquad k=\frac{\xi\uppi}{N}
\end{equation}
where $\xi=1,2,...N-1$ is the eigenmode number, sorted from the brightest (largest linewidth $=\Im\omega_{\xi}$) to the darkest (smallest linewidth).  The solution of Eq.~\eqref{eq:ans} can be recovered most easily in the limit $\varphi\ll k\ll \uppi$. In this case one can assume that $\cot \varphi\approx 1/\sin\varphi=1/\varphi$.
Neglecting the radiative decay, the Schr\"odinger equation Eq.~\eqref{eq:iS} yields the usual parabolic dispersion,
\begin{equation}
	\frac{{\Gamma_{10}}/2}{\omega_{\xi}^{(0)}}=-\frac{k^2}{2\varphi},\quad \psi_{\xi,s}=\sqrt{\frac{2}{N}}\cos k(s-\tfrac{1}{2})\:.
\end{equation}
Now we take into account the radiative decay  in Eq.~\eqref{eq:matrix} by considering the imaginary terms in the first order of the perturbation theory:
\begin{equation}\label{eq:order1}
	\frac{{\Gamma_{10}}/2}{\omega_{\xi}}=\frac{{\Gamma_{10}}/2}{\omega_{\xi}^{(0)}}-\frac{{\rm i}}{2}(| \psi_{\xi,1}|^2+| \psi_{\xi,N}|^2)=
	-\frac{\uppi^2\xi^2}{2\varphi N^2}-\frac{2{\rm i}}{N}\:,
\end{equation}
which is equivalent to Eq.~\eqref{eq:ans} in the considered limit of small $\xi\ll N$. Inverting Eq.~\eqref{eq:order1} we find for $\varphi\ll 1$
\begin{equation}
	\Gamma_\xi=2\text{Im}(\omega_\xi)=\Gamma_{10}\frac{N^3}{\xi^4}\frac{8 \varphi^2}{\uppi^4}\:.
\end{equation}

\section{Supplementary Note 5: Subradiant mode suppression due to the Fano-like interference}

\begin{figure}[htb!]
	\includegraphics[width=\columnwidth]{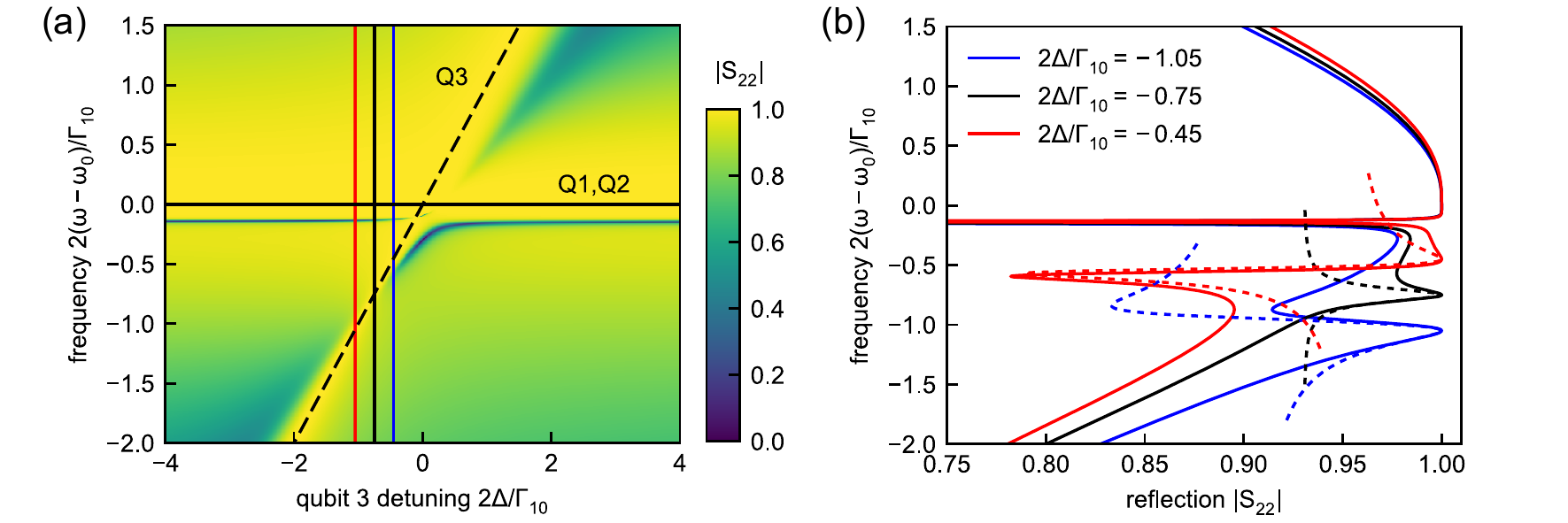}
	\caption{(a) Reflection spectra from an array of 3 qubits where the first two are tuned to the  the frequency $\omega_0$ (solid horizontal line) and last one to $\omega_0+\Delta$  (dashed inclined line) vs. the last qubit detuning $\Delta$. (b) Reflection spectra for three values of detuning, that are indicated on the graph,
		and shown by the vertical lines in (a). Solid curves present the results of  numerical calculation using Eq.~\eqref{eq:rt}, dashed curves have been obtained from the analytical Eq.~\eqref{eq:ran}.
		Calculation has been performed for $\varphi=\omega_0d/c=0.15$.}
	\label{fig:Fano}
\end{figure}
In this section we consider light reflection from the $N$-qubit array where first $N-1$ qubits are in resonance and the last
qubit is  detuned from the resonance. Our goal is to explain analytically the disappearance of the reflection dips for certain values of the detuning, for example at the frequency $7.894\,$GHz in Figure~3(a) of the main text.
In order to obtain a better understanding of the resonances in the reflection,  we use an approach based on the Hamiltonian  Eq.~\eqref{interactham}. In the case when  the inductances $L_{1,2}$  at the waveguide edges, leading to additional reflections, are not taken into account, this approach is exactly equivalent to the transfer matrix method~\cite{Ivchenko1994,Kosobukin2007}.
We start by solving the equation for the dimensionless dipole moments of the qubits $\psi_s$ induced by the incoming wave at the frequency $\omega$,
\begin{equation}\label{eq:system}
	\sum\limits_{s=1}^{N}[H_{rs}+(\omega_{s}-\omega) \delta_{rs}]\psi_s={\rm e}^{-{\rm i}(r-1)\varphi}\:,\quad r=1\ldots N\:,
\end{equation} 
where we assume that the qubits are located at the points $x=0,d,\ldots (N-1)d$.
After the dipole moments $\psi_s$ have been found from the solution of the system Eq.~\eqref{eq:system}, the amplitude reflection and transmission coefficients $r$ and $t$ are given by
\begin{equation}\label{eq:rt}
	r=-\frac{{\rm i}\Gamma_{10}}{2}\sum\limits_{s=1}^{N}\psi_s{\rm e}^{{\rm i}(s-1)\varphi},\quad 
	t=1-\frac{{\rm i}\Gamma_{10}}{2}\sum\limits_{s=1}^{N}\psi_s{\rm e}^{-{\rm i}(s-1)\varphi}\:.
\end{equation} 
In the Markovian approximation the phase  $\varphi=\omega d/c$ in Eqs.~\eqref{eq:system},\eqref{eq:rt} is evaluated at the qubit resonance frequency, $\varphi=\omega_0d/c$. Hence,  Eqs.~\eqref{eq:system},\eqref{eq:rt} reduce to a standard input-output problem.

We  will now illustrate the interferences in reflection for the specific case of $N=3$ qubits with the resonance frequencies $\omega_{0},\omega_{0}$ and $\omega_{0}+\Delta$, respectively. Our goal is to examine the light-induced  coupling  between the resonant dimer formed by the first two qubits with the last qubit and examine  the  Fano-like interferences in more detail.  Since we consider the situation when $\varphi\ll 1$, we describe the first  two resonant qubits by a symmetric superradiant state,
$
\psi_1=\psi_2=\frac{1}{\sqrt{2}}\psi_{\rm SR}\:.
$
As a result, the system Eq.~\eqref{eq:system} in the reduced basis reads
\begin{align}\label{eq:system2}
	\left(\omega_{0}+{\rm i} \Gamma_{10}-\omega\right)\psi_{\rm SR}&+\frac{{\rm i}\Gamma_{10}}{2\sqrt{2}}({\rm e}^{-{\rm i} \phi}+{\rm e}^{-2{\rm i} \phi})\psi_{3}=\frac{1+{\rm e}^{-{\rm i}\varphi}}{\sqrt{2}}\:,\\
	\left(\omega_{0}+\Delta+\frac{{\rm i} \Gamma_{10}}{2}-\omega\right)\psi_{3}&+\frac{{\rm i}\Gamma_{10}}{2\sqrt{2}}({\rm e}^{-{\rm i} \phi}+{\rm e}^{-2{\rm i} \phi})\psi_{\rm SR}={\rm e}^{-2{\rm i}\varphi}\:.\nonumber
\end{align}
and the reflection coefficient is given by
\begin{equation}\label{eq:r2}
	r=-\frac{{\rm i}\Gamma_{10}}{2}\left(\frac{1+{\rm e}^{-{\rm i}\varphi}}{\sqrt{2}}\psi_{\rm SR}+{\rm e}^{-2{\rm i}\phi}\psi_{3} \right)\:.
\end{equation}
Now we restrict ourselves to the frequency range where the frequency is close to the detuned qubit resonance, i.e.
\begin{equation}
	|\omega-\omega_0-\Delta|\ll \Delta\:.
\end{equation}
In this spectral range the reflection coefficient Eq.~\eqref{eq:r2}, obtained from the system Eq.~\eqref{eq:system2}, can be approximately presented as
\begin{equation}\label{eq:ran}
	r\approx {\rm i} r_0 \frac{\omega-\omega_0-\Delta+\frac{\Gamma_{10}}{2}(1/r_0^*-{\rm i})}{\omega-\omega_0-\Delta+\frac{\Gamma_{10}}{2}(r_0-{\rm i})}
\end{equation}
where
\begin{equation}\label{eq:r0}
	r_0 = \frac{1}{-{\rm i} +\frac{1}{3}(2\Delta/\Gamma_{10}+5\varphi)+\frac{4{\rm i} \varphi}{3}\Delta/\Gamma_{10}}\:,
\end{equation}
and we assume that $\varphi\ll 1$.
Here, Eq.~\eqref{eq:r0} describes the slow varying background of the reflection coefficient Eq.~\eqref{eq:ran}. This background corresponds to the mode, where the third qubit oscillates in phase with the first two. This interpretation becomes  most clear in the regime where all the qubits are in the same point, $\varphi=0$,
so that
\begin{equation}\label{eq:r0SR}
	r_0=\frac{3}{-3{\rm i} +2\Delta/\Gamma_{10}}\:.
\end{equation}
Equation Eq.~\eqref{eq:r0SR} describes just the resonant reflection determined by the superradiant mode of   3 qubits~\cite{Ivchenko1994,chang_cavity_2012}. The second factor in the reflection coefficient Eq.~\eqref{eq:ran} describes the resonant coupling of the last qubit with the superradiant mode. This factor has a resonance at the frequency $\omega_0+\Delta-\Re r_0\Gamma_{10}/2$ with the radiative decay rate $(1-\Im r_0)\Gamma_{10}/2$. Both the radiative decay and the position of the resonance depend on the phase of the background reflection Eq.~\eqref{eq:r0SR} at the resonance frequency of the detuned qubit $\omega_0+\Delta$. 
This is very similar to the general picture of Fano interference between two scattering channels with broad and narrow  spectral resonances, resulting in characteristic asymmetric spectral lines~\cite{Limonov2017}.
The reflection coefficient Eq.~\eqref{eq:r0} cannot be completely reduced to the Fano equation because, contrary to the Fano case, both the superradiant mode and the last qubit mode are  directly coupled to the input and output channels in Eq.~\eqref{eq:system2}. However, Eq.~\eqref{eq:r0} also yields asymmetric reflection spectra, as is demonstrated by the calculation in Supplementary Figure~\ref{fig:Fano}. In this figure, similarly to Figure~3 of the  main text, we show the numerically calculated reflection spectra depending on the detuning $\Delta$. Right panel presents the spectra for three values of the detuning (solid lines) compared with the analytical result Eq.~\eqref{eq:ran} (dashed lines).  Similarly to Figure~3 of the  main text, Supplementary Figure~\ref{fig:Fano} has a blindspot for the detuning 
$2\Delta/\Gamma_{10}=-0.75=-5\varphi$ [vertical black line in Supplementary Figure~\ref{fig:Fano}(a)]. The calculation in Supplementary Figure~\ref{fig:Fano}(b) demonstrates that the exact result is well described by the approximation Eq.~\eqref{eq:ran} in the vicinity of the resonance of the detuned qubit. It can be seen, that for $2\Delta/\Gamma_{10}+5\varphi=0$, the background reflection coefficient Eq.~\eqref{eq:r0} becomes purely imaginary. As such, the background provides a constructively interfering contribution at the last qubit resonance and results in the symmetric reflection peak. This  is demonstrated by the  black curves in Supplementary Figure~\ref{fig:Fano}(b). When the last qubit frequency $\omega_0+\Delta$ is detuned from $\omega_0-(5/2)\Gamma_{10}\varphi$, the interference stops being  constructive, resulting in the asymmetric reflection resonances [blue  and red lines in Supplementary Figure~\ref{fig:Fano}(b)].
The two-mode model Eq.~\eqref{eq:system2}, describing  Fano-like interferences between the modes of the last qubit and the given mode 
of first $N-1$ resonant qubits, has a very general character. It  can be generalized for arbitrary values of  $N$, explaining the blind spots in Figure~3 of the main text.

\section{Supplementary Note 6: Transfer matrix approach to Autler-Townes splitting}
In order to calculate the transmission of the collective Autler-Townes splitting of the qubit array, we use the transfer matrix approach as described above, however based on the reflection of a dressed 3-level system as derived in \cite{abdumalikov_electromagnetically_2010}:
\begin{equation}
	r^\text{3L}=-\frac{\Gamma_{10}}{2[\gamma_{10}+{\rm i}(\omega-\omega_{10})]+\frac{\Omega_\text{c}^2}{2\gamma_{20}+2{\rm i}(\omega-\omega_{10}+\omega_\text{c}-\omega_{21})}}
	\label{3L}
\end{equation}
Here, a second driving control field with frequency $\omega_\text{c}$ and Rabi strength $\Omega_\text{c}$ is included. The probe field is assumed to be weak ($\Omega_{\rm p}\ll \gamma_{10}$). The frequency of the $1\rightarrow 2 $ transition 
is $\omega_{21}$ and $\gamma_{20}$ is the decoherence rate of the $0\rightarrow 2 $ transition.  We note that the transfer matrix approach does not include the scattering and interference effects of the control tone in the array.\\

\begin{figure}
	\includegraphics[width=\columnwidth]{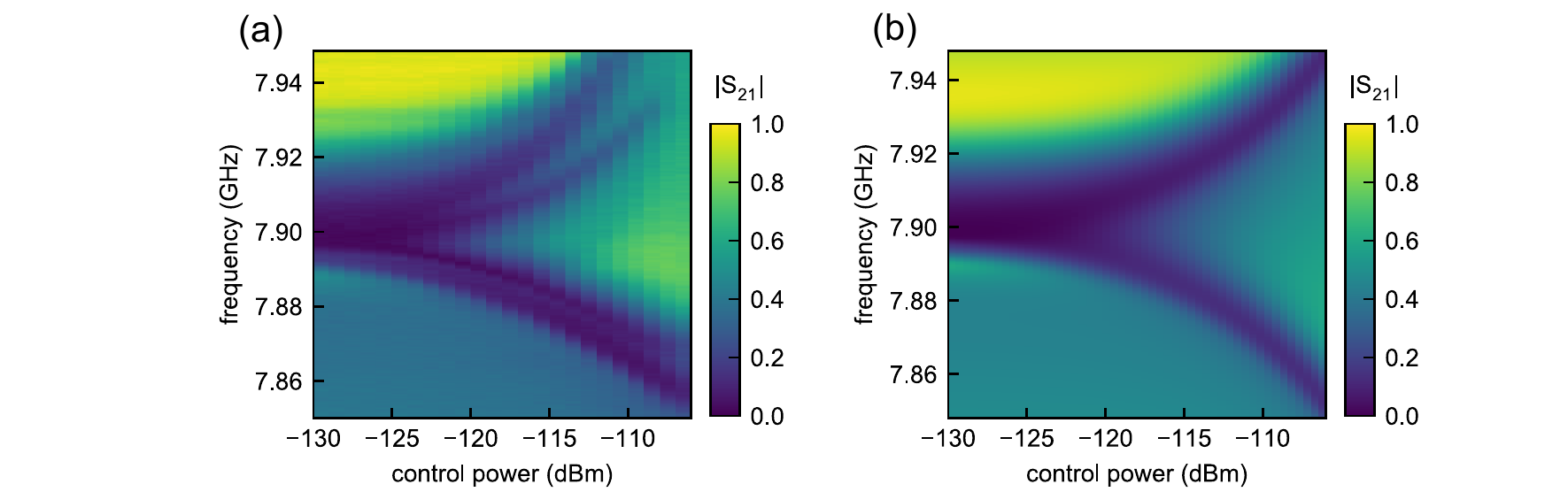}
	\caption{(a) Measured transmission of a collective Autler-Townes splitting with $N=8$ resonant qubits. (b) Calculated Autler-Townes splitting based on the transfer matrix. (Used parameters: $\frac{\omega_{10}}{2\uppi}=7.898\,\text{GHz},\, \frac{\omega_{{\rm c}}}{2\uppi}=\braket{\frac{\omega_{21}}{2\uppi}}=7.623\,\text{GHz},\,\Gamma_{10}/2\uppi=6.4\,\text{MHz},\, \gamma_{10}/2\uppi=3.4\,\text{MHz},\, \gamma_{20}/2\uppi=11.1\,$MHz)
	}
	\label{ats_comparison_8q}
\end{figure}
Supplementary Figure~\ref{ats_comparison_8q} shows the comparison between the measured 8 qubit ATS as presented in the main text and the transfer matrix result with included cable resonance. All model parameters used in the calculation were extracted from the fits shown in Figure~1(b) in the main text and fits to $t^\text{3L}(\Omega_{\rm c})=1+r^\text{3L}$ for the case of a single qubit and a resonant pump and control tone. Small differences between simulation and calculation can be found in the fine-structure of the two branches of the ATS. Their origin lies in small  imperfections of the qubit anharmonicity, or a slight detuning from the common resonance frequency.

\end{document}